\RequirePackage{fix-cm}
\documentclass[twocolumn,epjc3]{svjour3}  
\smartqed  
\RequirePackage{graphicx}
\usepackage{amsmath,amssymb,aas_macros}
\usepackage{multirow}
\usepackage[mathlines]{lineno}
\RequirePackage[colorlinks,citecolor=blue,urlcolor=blue,linkcolor=blue]{hyperref}
\journalname{Eur. Phys. J. C}
\begin{document}

\title{Search for dark matter from the center of the Earth with ten years of IceCube data
}



\author{First Author\thanksref{e1,addr1}
        \and
        Second Author\thanksref{e2,addr2,addr3}
}
\onecolumn
\author{R. Abbasi\thanksref{loyola}
\and M. Ackermann\thanksref{zeuthen}
\and J. Adams\thanksref{christchurch}
\and S. K. Agarwalla\thanksref{madisonpac,a}
\and J. A. Aguilar\thanksref{brusselslibre}
\and M. Ahlers\thanksref{copenhagen}
\and J.M. Alameddine\thanksref{dortmund}
\and N. M. Amin\thanksref{bartol}
\and K. Andeen\thanksref{marquette}
\and C. Arg{\"u}elles\thanksref{harvard}
% ...
}
\authorrunning{IceCube Collaboration}
\thankstext{a}{also at Institute of Physics, Sachivalaya Marg, Sainik School Post, Bhubaneswar 751005, India}
\thankstext{b}{also at Department of Space, Earth and Environment, Chalmers University of Technology, 412 96 Gothenburg, Sweden}
\thankstext{c}{also at Earthquake Research Institute, University of Tokyo, Bunkyo, Tokyo 113-0032, Japan}
\institute{
% ...
Universit{\'e} Libre de Bruxelles, Science Faculty CP230, B-1050 Brussels, Belgium \label{brusselslibre}
% ...
\and Department of Physics and Laboratory for Particle Physics and Cosmology, Harvard University, Cambridge, MA 02138, USA \label{harvard}
% ...
\and Department of Physics, Loyola University Chicago, Chicago, IL 60660, USA \label{loyola}
\and Dept. of Physics and Astronomy, University of Canterbury, Private Bag 4800, Christchurch, New Zealand \label{christchurch}
% ...
\and Niels Bohr Institute, University of Copenhagen, DK-2100 Copenhagen, Denmark \label{copenhagen}
\and Dept. of Physics, TU Dortmund University, D-44221 Dortmund, Germany \label{dortmund}
% ...
\and Dept. of Physics and Wisconsin IceCube Particle Astrophysics Center, University of Wisconsin{\textemdash}Madison, Madison, WI 53706, USA \label{madisonpac}
% ...
\and Department of Physics, Marquette University, Milwaukee, WI 53201, USA \label{marquette}
% ...
\and Bartol Research Institute and Dept. of Physics and Astronomy, University of Delaware, Newark, DE 19716, USA \label{bartol}
% ...
\and Deutsches Elektronen-Synchrotron DESY, Platanenallee 6, D-15738 Zeuthen, Germany \label{zeuthen}
}
\date{Received: date / Accepted: date}
\maketitle
\twocolumn

\begin{abstract}
The nature of dark matter remains unresolved in fundamental physics. Weakly Interacting Massive Particles (WIMPs), which could explain the nature of dark matter, can be captured by celestial bodies like the Sun or Earth, leading to enhanced self-annihilation into Standard Model particles including neutrinos detectable by neutrino telescopes such as the IceCube Neutrino Observatory.  

\begin{sloppypar}
This article presents a search for muon neutrinos from the center of the Earth performed with 10 years of IceCube data using a track-like event selection. We considered a number of WIMP annihilation channels ($\chi\chi\rightarrow\tau^+\tau^-$/$W^+W^-$/$b\bar{b}$) and masses ranging from 10 GeV to 10 TeV. No significant excess over background due to a dark matter signal was found while the most significant result corresponds to the annihilation
channel $\chi\chi\rightarrow b\bar{b}$ for the mass $m_{\chi}=250$~GeV with a post-trial significance of $1.06\sigma$. Our results are competitive with previous such searches and direct detection experiments. Our upper limits on the spin-independent WIMP scattering are world-leading among neutrino telescopes for WIMP masses $m_{\chi}>100$~GeV.   
\end{sloppypar}
\end{abstract}

\section{Introduction}
\label{sec:intro}
Over the last century, an increasing amount of evidence has emerged indicating that approximately $22\%$ of the Universe and about $85\%$ of its matter content~\cite{planckI:2018} is composed of unknown matter~\cite{introDM}. This elusive form of matter, referred to as {\it dark matter} (DM), must exhibit a low probability of interaction with ordinary matter if it is considered to be of corpuscular nature. 

The existence of DM is supported by various observations and experiments. These include measurements of cosmological parameters conducted by the Planck collaboration~\cite{planckI:2018}, as well as estimations of the mass content of galaxies and galaxy clusters through gravitational lensing or velocity dispersion measurements~\cite{zwicky1933rotverschiebung,bahcall1995dark,bahcall2014tracing,Clowe:2006eq}.

Numerous theoretical models propose the existence of dark matter particles beyond the {\it Standard Model} (SM) of particle physics. One set of proposed particles, collectively called {\it Weakly Interacting Massive Particles} (WIMPs), interact only gravitationally and by forces as weak as, or weaker than, the Weak Interaction. As an example, supersymmetric extensions of the SM predict the existence of a stable dark matter particle with these properties, the neutralino~\cite{DM_rev:96}. However, searches for these WIMPs have, so far, yielded no results, while collider experiments have failed to find hints of supersymmetry~\cite{Adam_2022}. 

This article is structured as follows: in Sec.~\ref{sec:1} we explain the neutrino production from dark matter self-annihilation at the center of Earth as well as the capturing of dark matter by celestial bodies. Section~\ref{sec:icecube} describes the IceCube neutrino telescope as well as the signal and background simulation. The event selection is explained in Sec.~\ref{sec:event_selection}. The analysis method and results are described in Sec.~\ref{sec:method} and Sec.~\ref{sec:results} respectively. Results are discussed in Sec.~\ref{sec:discu} and finally, conclusions are given in Sec.~\ref{sec:conclu}.

\section{Neutrinos from dark matter self-annihilation at the center of the Earth}
\label{sec:1}
\begin{sloppypar}
The Milky Way is assumed to be embedded in a dark matter halo~\cite{rubin1970rotation,roberts1973comparison,rubin1980rotational}. As the Solar System travels through this halo, DM particles can be captured by the gravitational potential of celestial bodies, such as the Sun or the Earth. This capture occurs through repeated scattering interactions between these bodies' nuclei and DM particles in intersecting orbits. As a result of these interactions, dark matter particles can lose kinetic energy resulting in velocities below the escape velocity of the celestial object, and accumulate in the central regions of the celestial body~\cite{1987ApJ...321..571G,PhysRevLett.81.5726}. Dark matter in these over-densities can subsequently undergo self-annihilation processes leading to the production of SM particles. This annihilation process depletes the central regions of DM reducing the dark matter number density, $N_\chi$. The competing processes of capture and self-annihilation can be described by the following equation:   
\end{sloppypar}

\begin{equation}
    \label{eq:dm_rate}
    \dot{N_{\chi}} = C_{\textrm{C}}-C_{\textrm{A}}N_{\chi}^2-C_{\textrm{E}}N_{\chi},
\end{equation}

\noindent where $C_{\textrm{C}}$ is the capture rate which depends on the local DM density, the scattering cross section, and the chemical composition of the target body, and $C_\textrm{A}N_{\chi}^2=\Gamma_{\textrm{A}}$ is the annihilation rate which is proportional to $N_\chi^2$ since two dark matter particles are required for annihilation. For thermalized dark matter, the parameter $C_\textrm{A}$ is proportional to the self-annihilation cross section averaged over velocity, $\langle\sigma_{\textrm{A}}v\rangle$ --with $v$ being the relative velocity between the two dark matter particles--, as well as the Earth effective volume which depends on the density and temperature at the center of the Earth~\cite{Baratella_2014}. The last term, $C_\textrm{E} N_{\chi}$, corresponds to the rate of {\it evaporation}, i.e. when DM gains speed above the escape velocity in scattering off ambient nuclei, thus escaping from the capturing celestial body. This rate is negligible for the DM masses ($> 10\; \textrm{GeV}$) considered in this work ~\cite{evaporation:2021,Busoni_2013}. Hence Eq.~\ref{eq:dm_rate} has an equilibrium (corresponding to $\dot{N}_\chi = 0$):
 
\begin{equation}
    \label{eq:dm_rate_sol}
    \Gamma_{\textrm {A}}(t) = \frac{1}{2}C_{\textrm{C}}\tanh^2\left(\frac{t}{\tau}\right),
\end{equation}

\noindent where $\tau=(C_{\textrm{C}}C_{\textrm {A}})^{-1/2}$ is the characteristic time scale for the capture and annihilation processes to come into equilibrium. For times $t \gg \tau$ Eq.~\eqref{eq:dm_rate_sol} becomes $C_{\textrm{C} }= 2\Gamma_{\textrm{A}}$ and the annihilation rate reaches its maximal value, i.e. when two DM particles are captured, two are annihilated. However, given the age of the Earth, $t_\oplus = 4.5 $ Gyr, and canonical values of scattering and annihilation cross section~\cite{Kolb:1990vq}, it is safe to assume that DM inside the Earth cannot have reached equilibrium~\cite{GREEN2019120}. In this situation, in order to solve Eq.~\ref{eq:dm_rate_sol} one needs to adopt a specific value of the annihilation cross section, $\langle\sigma_{\textrm{A}}v\rangle$, in order to derive a constraint on the scattering cross section. 

\begin{figure}
    \includegraphics[width=.49\textwidth]{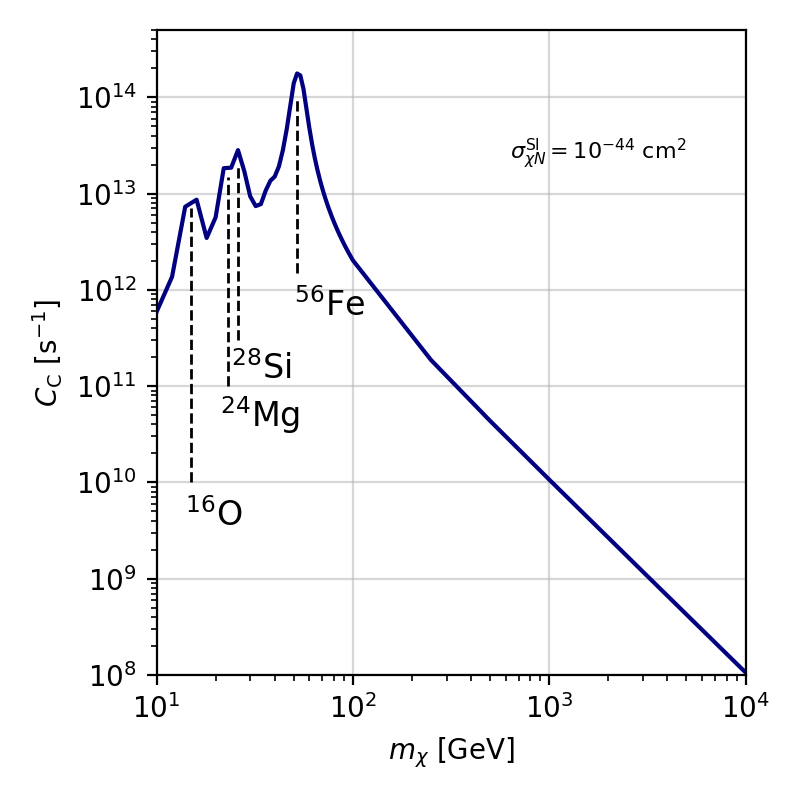}
    \caption{Capture rate as a function of the dark matter particle mass for the spin-independent WIMP-nucleon scattering cross section value \mbox{$\sigma_{\chi N}^{\textrm {SI}}=10^{-44}\;\textrm{cm}^2$}. The peaks correspond to resonance capture with the most abundant elements on Earth: O, Mg, Si, and Fe. Computed using DarkSusy \cite{DarkSusy6,DarkSusyI,DarkSusy:code}.}
    \label{fig:cap_rate}
\end{figure}

\paragraph{Capturing.--- }
The Earth is mainly composed of heavy elements, hence the most dominant scattering process is the spin-independent DM-nucleon scattering, which is governed by the spin-independent DM-nucleon scattering cross section $\sigma_{\chi N}^{\textrm{SI}}$. The total capture rate can be expressed as a the sum of capture rates on each of the species of nuclei in the Earth, $N_i$~\cite{DM_rev:96}:

\begin{multline}
        \label{eq:C_C}
        C_{\textrm{C}}
        =c\frac{\rho_{\chi}}{m_{\chi}\bar{v}}\sum_iF_i(m_{\chi})\sigma_{\chi N_i}^{\textrm{SI}}f_i\phi_i\frac{S(m_{\chi}/m_{N_i})}{m_{N_i}},
    \end{multline}

where $\rho_{\chi}/m_\chi$ is local dark matter number density and $\bar{v}$ the velocity dispersion of dark matter. The terms $\sigma_{\chi N_{i}}^{\textrm{SI}}$, $f_i$, and $\phi_i$ are, respectively, the DM-nucleon cross section, the fractional abundance, and the dimensionless gravitational potential of the $i$-th nuclear element. $F_i(m_{\chi})$ is the Helm form factor encoding the coherent scattering of DM off multiple nucleons in the nucleus~\cite{1987ApJ...321..571G}. Finally, the term $S(m_{\chi}/m_{N_i})$ is the kinematic suppression factor for a dark matter particle of mass $m_\chi$ scattering off a nucleus $i$ of mass $m_{N_i}$. This term tends to unity when the DM mass is close to $m_{N_i}$. This kinematic suppression factor leads to resonance features in the capture rate when the dark matter mass coincides with one of the elements of the Earth composition as can be seen in Fig.~\ref{fig:cap_rate} which shows the capture rate on Earth as a function of the dark matter mass.


\paragraph{Annihilation.--- }
Neutrinos are among both direct and secondary products from DM self-annihilation, and in the case of annihilation at the center of the Earth, the only particles that can reach its surface. The expected flux of neutrinos will depend on the annihilation rate and the branching fractions of the different annihilation channels. In particular the muon neutrino volumetric flux (or event rate induced by muon neutrinos per unit volume) can be obtained via the relation~\cite{DM_rev:96}:
    
    \begin{multline}
        \label{eq:vol_flux2ann_rate}
        \Gamma_{\nu\rightarrow\mu} =
        \frac{\Gamma_{\rm A}}{4\pi R^2}\times\\
        \times\int_{0}^{\infty}dE_{\nu_\mu}\sigma_{\nu_\mu N}(E_{\mu}|E_{\nu_\mu})\rho_N\sum_j \mathcal{B}_j \left(\frac{dN_{\nu_\mu}}{dE_{\nu_\mu}}\right)^{j},
    \end{multline}
    
\noindent where $R$ corresponds to the Earth radius, $\sigma_{\nu_{\mu} N}(E_{\mu}|E_{\nu_\mu})$ is the integrated neutrino-nucleon cross section over the muon energy for muon neutrinos with energy $E_{\nu_{\mu}}$ producing a muon with energy $E_\mu$, $\rho_N$ is the nucleon density at the detector, $\mathcal{B}_j$ are the branching fractions, and $\left(dN_{\nu_{\mu}}/dE_{\nu_{\mu}}\right)^{j}$ are the muon neutrino spectra for the $j$th DM self-annihilation channel. The Earth is small enough that absorption of neutrinos as they pass through the Earth can be neglected at energies $< 30$ TeV~\cite{DM_rev:96}. As can be seen in Eq.~\eqref{eq:vol_flux2ann_rate}, measuring the neutrino flux corresponds to a measurement of the annihilation rate and by using Eq.~\eqref{eq:dm_rate_sol} we can relate it to the capture rate once a value of the annihilation cross section is assumed. This is unlike the case of the Sun where equilibrium is reached~\cite{DM_rev:96}. Typically  the canonical value of $\langle\sigma_{\textrm{A}}v\rangle = 3\times 10^{-26}\textrm{cm}^3~\textrm{s}^{-1}$ for a WIMP that makes up all the DM~\cite{Kolb:1990vq} is used for calculating limits, however a more general way is to present them in the $\langle\sigma_{\textrm{A}}v\rangle - \sigma_{\chi N}^{\textrm{SI}}$ plane (see Sec.~\ref{sec:results} for more details).

Limits on the neutrino flux coming from the center of the Earth, have been published over the last twenty years by experiments such as ANTARES \cite{ANTARES_Earth:2016}, SuperKamiokande \cite{SuperK_DM:2020}, and AMANDA~\cite{AMANDA_Earth}. A previous search was performed in IceCube~\cite{EarthIce:2016} using only one year of data. In this study we extend the search to ten years of data with a re-optimized event selection. We use the zenith angle of the direction of the neutrino event as an observable, and, unlike in our previous analysis, the energy of the event is also now considered. 

\section{The IceCube Neutrino Observatory}
\label{sec:icecube}
\begin{sloppypar}   
The IceCube Neutrino Observatory~\cite{IceCube:2016a} is a one-cubic-kilometer neutrino telescope situated at the geographic South Pole. The detector is composed of 5,160 photomultiplier tubes housed in individual detector units called Digital Optical Modules (DOMs)~\cite{Abbasi_2009} deployed on 86 strings in the Antarctic ice-cap between the depths of 1450 m and 2450 m. IceCube detects Cherenkov light induced by the passage of superluminal charged leptons and hadrons produced in neutrino interactions with the surrounding ice. The number of detected Cherenkov photons and their arrival time are used to reconstruct the direction and energy of the incoming neutrino. Depending on the neutrino interaction, different event morphologies can be observed. The muon resulting from $\nu_\mu$ interacting via the charge current interaction will leave a track-like event providing directional information, while neutral-current interactions of all flavors and charge-current interactions of $\nu_e$ and $\nu_\tau$ will induce a hadronic or electromagnetic shower resulting in a cascade-like event. IceCube has detected neutrinos with energies ranging from a few GeV to a few PeV. A denser sub-array called {DeepCore}~\cite{DEEPCORE2012615}, is placed between 2100~m and 2450~m of depth and is dedicated to the detection of neutrinos between 1~GeV and 100~GeV. A surface detector called {IceTop}~\cite{ICETOP2013188} is used for cosmic-ray studies as well as to veto atmospheric muons in the in-ice detector.
\end{sloppypar}


Neutrinos from dark matter self-annihilation at the center of the Earth are expected to reach the detector with highly vertical up-going directions. This direction corresponds to zenith angles of $\theta \sim 180^\circ$ and represents a unique direction in local coordinates, making it impossible to estimate the background from an \textit{off-source} region in the sky, as there is no equivalent direction in local coordinates from which we can estimate the background. 
Right ascension scrambling, a technique typically used in neutrino telescopes to preserve the background's declination dependence and detector efficiency~\cite{Braun_2010}, is not feasible for this analysis because at the South Pole, declination and zenith angles are complementary, and right ascension scrambling will not dilute any possible signal from the center of the Earth. 
For this reason, we must rely on Monte Carlo simulations to model the background, for the optimization of the event selection, and for the statistical analysis. The different sources of background for this analysis are muons and neutrinos generated in cosmic-ray interactions in the atmosphere (atmospheric muons/neutrinos) and, to a lesser extent, astrophysical neutrinos.
Although atmospheric muons are only down-going, a fraction of them are mis-reconstructed as up-going making these muons the majority of background events at the initial level of the event selection. At the final level, the atmospheric neutrino contribution remains as the most abundant background component with a rate of $3\times10^{-5}\textrm{Hz}$. 

\paragraph{Dataset.--- }
We used 3619 days of data taken over ten detector seasons, from May 2011 to May 2020. A subset of 353 days of data, taken sparsely over the ten seasons, was used as a verification dataset during the event selection development and has been omitted from the final analysis, leaving a total of 3266 days of data.
\paragraph{Background simulation.---}
Atmospheric muons are simulated with the \textsc{Corsika} package~\cite{corsika:98}. Neutrinos, both atmospheric and astrophysical, are simulated with \textsc{Genie}~\cite{GENIEAndreopoulos:2015wxa} for energies up to 100 GeV and with \textsc{NuGen}~\cite{ANIS_GAZIZOV:2005203} for energies above 100 GeV. \textsc{Genie} includes a more complete implementation of the various neutrino interactions at low energies including Quasi-Elastic, Resonance, and \textit{Deep Inelastic Scattering} (DIS) cross sections~\cite{GENIEAndreopoulos:2015wxa}. \textsc{NuGen}, however implements only the DIS neutrino cross section based on the~\textit{CSMS} calculation~\cite{Cooper-Sarkar:2011jtt} as this is the dominant process at high energies. Neutrino interactions produce secondary charge particles that propagate through the detector inducing Cherenkov light. The Monte Carlo also simulates the propagation of the Cherenkov light in ice, and its detection by the IceCube DOMs. These simulated neutrinos are used to estimate atmospheric (including the prompt~\cite{Sarcevic:PhysRevD.78.043005}) and astrophysical~\cite{IceCubeDiffuse:2021uhz} backgrounds.

\paragraph{Signal simulation.---}
\begin{sloppypar}
Neutrinos from self-annihilation of dark matter in the center of the Earth are simulated via \textsc{WimpSim}~\cite{WimpSim:2008,WimpSim:code}. \textsc{WimpSim} uses \textsc{PYTHIA}~\cite{Pythia} to simulate the hadronization and decay of the annihilation products that produce neutrinos of all flavors. \textsc{WimpSim} also propagates the neutrino on its way to the surface taking into account the neutrino oscillations. As with atmospheric and astrophysical neutrinos, they are propagated through the detector simulating Cherenkov light production, propagation, and detection by IceCube DOMs. We simulated neutrinos produced in self-annihilation of DM with mass ranging from 10~GeV to 10~TeV in three main annihilation channels: \mbox{$\chi\chi\rightarrow\tau^+\tau^-$/$W^+W^-$/$b\bar{b}$}. Annihilation to $b\bar{b}$ produces a soft neutrino spectrum, while $W^+W^-$ produces a hard spectrum. Annihilation to $\tau^+\tau^-$ replaces the $W^+W^-$ channel for DM lighter than the $W$ boson. Details of the different mass-channel combinations which were used in this analysis are listed in Tab.~\ref{table:wimpsim}. Since this analysis uses the direction of the neutrinos as an observable, only muon neutrinos are considered for these simulations.
\end{sloppypar} 
\begin{table}
    \caption{Summary of WIMPs simulation scenarios produced with \textsc{WimpSim}.}
    \label{table:wimpsim}
    \begin{tabular}{c|c}
        Channel & Masses \\
        \hline
        \multirow{3}{*}{$\chi\chi\rightarrow\tau^+\tau^-$} & [10, 20, 35, 50] GeV\\
         & [100, 250, 500] GeV\\
         & [1, 3, 5, 10] TeV\\
         \hline
         \multirow{2}{*}{$\chi\chi\rightarrow W^+W^-$} & [100, 250, 500] GeV\\
         & [1, 3, 5, 10] TeV\\
         \hline
         \multirow{3}{*}{$\chi\chi\rightarrow b\bar{b}$} & [35, 50] GeV\\
         & [100, 250, 500] GeV\\
         & [1, 3, 5, 10] TeV\\
    \end{tabular}
\end{table}

\section{The event selection}
\label{sec:event_selection}

\begin{figure*}[ht]
    \begin{minipage}{0.49\textwidth}
    \includegraphics[width=\textwidth]{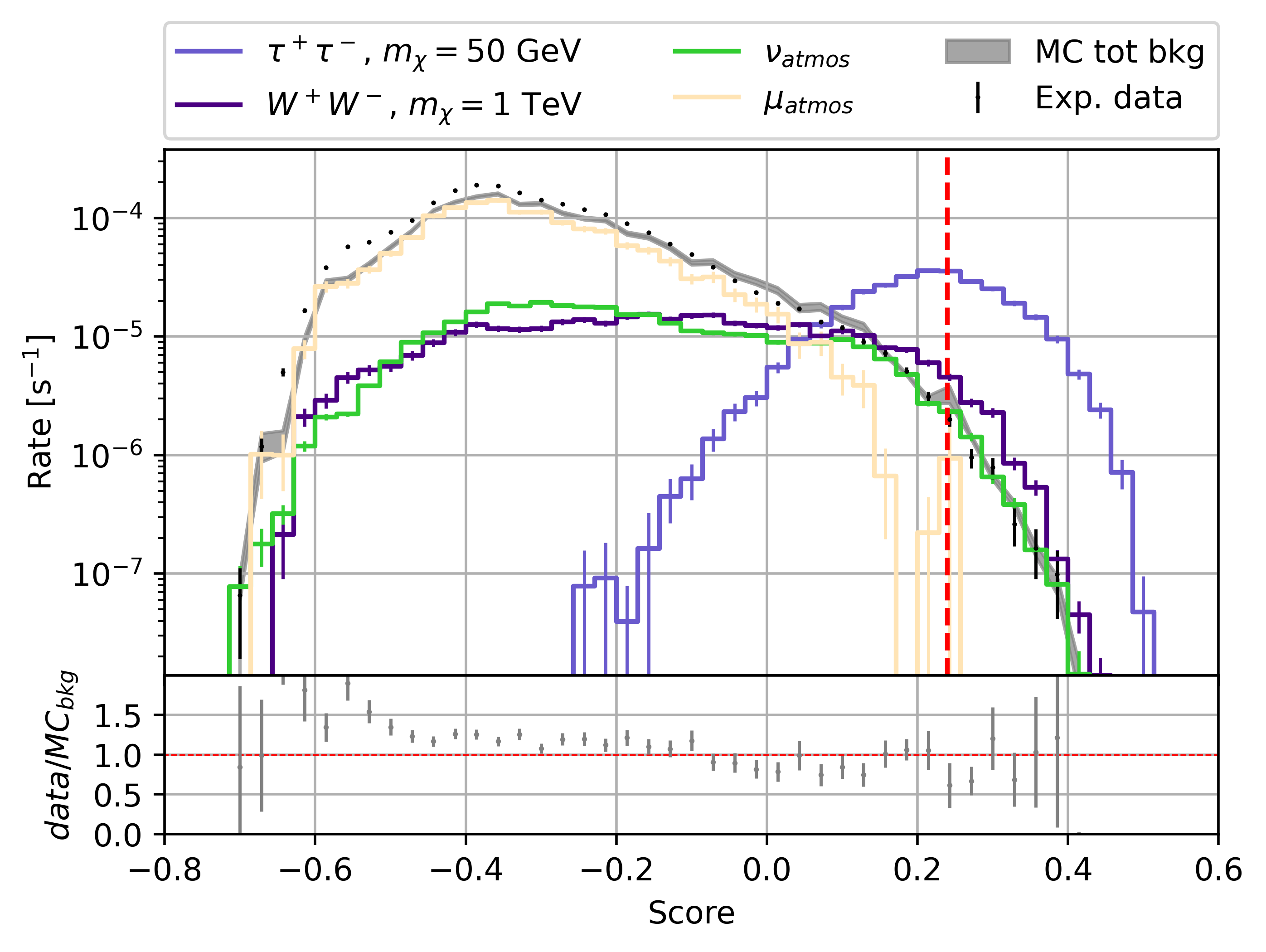}
    \end{minipage}
    \begin{minipage}{0.49\textwidth}
    \includegraphics[width=\textwidth]{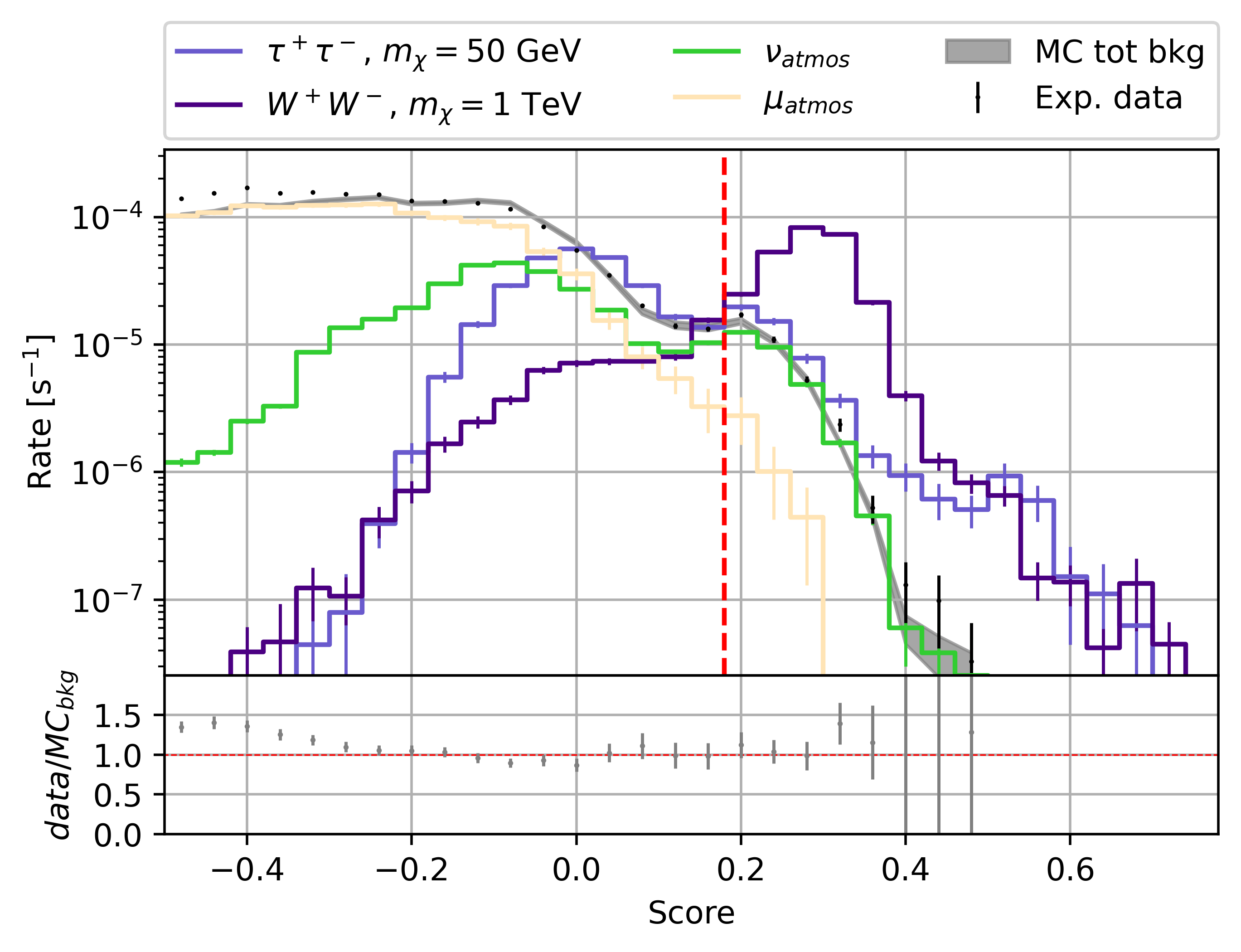}
    \end{minipage}
    \caption{Score distributions for the LE (left) and HE (right) BDTs. The baseline LE and HE signal configurations are in light purple and dark purple, respectively. Atmospheric neutrinos are shown in green and muons in light brown lines. The grey band indicates the total estimated background while the black dots represent the data verification sample.}
    \label{fig:bdt}
\end{figure*}

The event selection in this analysis is designed to create a dataset consisting of almost vertical up-going track-like events, while preserving the agreement between data and simulation. 
The selections starts by using a combination of general track and low energy event filters, together with an event filter aiming at selecting vertical tracks. As this analysis covers low energy events, it also makes use of DeepCore-specific filters~\cite{DEEPCORE2012615}. The next step in the event selection is called level 1 (\textit{L1}) and is designed to eliminate obvious background events, such as down-going tracks or muon tracks which are not fully contained within the detector. To this end, it cuts on the zenith and goodness-of-fit of reconstructed tracks. In addition, we cut out events with the reconstructed interaction vertex outside the detector or at the top of the detector volume. After this, we apply a cleaning algorithm that removes isolated hits which are not causally-connected (mainly due to the dark count rate from the photomultiplier tubes~\cite{PMT2010139}) in a step called level 2 (\textit{L2}). After hit cleaning, the track reconstructions are re-run and additional variables that help discriminate up- vs. down-going events are calculated~\cite{EarthIce:2016}. The same selection on the reconstructed zenith and quality of the reconstruction which were applied at \textit{L1} are re-applied at this stage called level 3 (\textit{L3}). For a more detailed description of these processing levels see~\cite{Renzi:2022}.
\begin{sloppypar}
This preliminary selection, from \textit{L1} to \textit{L3}, prepares the dataset for the next step, level 4 (\textit{L4}), which makes use of a boosted decision tree (BDT)~\cite{cite-key} for background discrimination. The software \texttt{pybdt}~\cite{Richman:2015}, developed within the IceCube collaboration, was used to this end. The BDT is trained to discriminate between background events and signal events. Once trained the BDT assigns a score to each event ranging from -1 (most background-like) to +1 (most signal-like). As this analysis probes a wide range of masses, signal originating from different dark matter masses will leave different event morphologies in the detector. In particular, low-energy events (when $m_\chi \lesssim 100\;\textrm{GeV}$) will not leave clear track-like signatures, and will be predominantly contained within the DeepCore volume. The direction of these low-energy events is poorly reconstructed due to the smaller number of photons detected. High-energy events ($m_\chi \gtrsim 100\;\textrm{GeV}$), on the other hand, can leave kilometer-long tracks in the IceCube volume. Given these differences, the event selection is split into two: a \textit{low-energy} (LE) selection trained on neutrino events from the annihilation of DM with $m_\chi=50\;\textrm{GeV}$, into $\tau^+\tau^-$, and a \textit{high-energy} (HE) selection trained on neutrino events from a higher DM mass, $m_\chi=1\;\textrm{TeV}$, annihilating into $W^+W^-$. Different observables and reconstructed parameters based on the containment, vertex position, and direction of the events were used to train the BDTs (for the descriptive list of variables used in the training please see~\cite{Renzi:2022}). The score distributions for the LE and HE BDTs are shown in Fig.~\ref{fig:bdt} illustrating the differences in the score distributions for the LE and HE baseline signal configurations across the two BDTs.

\end{sloppypar}
\begin{sloppypar}
    
\begin{figure}[ht]
    \includegraphics[width=0.49\textwidth]{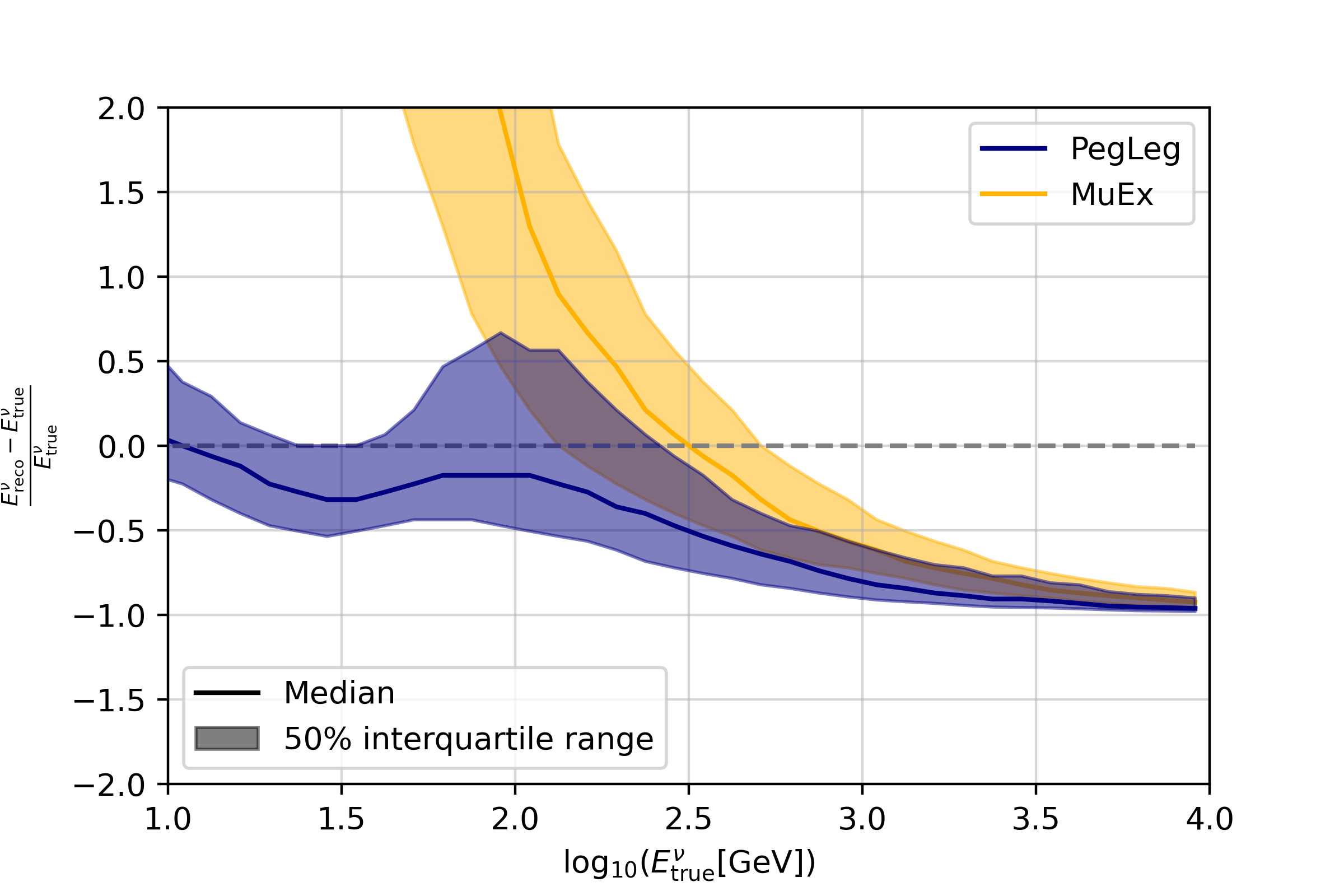}
    \caption{Comparison between the energy resolution for the final reconstruction algorithm \textsc{PegLeg} (blue) and the older reconstruction \textsc{MuEx} (yellow).}
    \label{fig:en_res}
\end{figure}

Once the BDT has been trained we can apply some nominal cuts excluding the most distinct (score close to -1) background events. After reducing the event selection rate by eliminating a large contribution of background events, we can perform a more sophisticated (and more computationally intense) energy reconstruction algorithm called $\textsc{PegLeg}$~\cite{Leuermann:2018oxl}. This energy reconstruction improves the overall energy resolution by $\sim11\%$ in terms of $\log_{10} E^\nu$ with respect to an older algorithm ($\textsc{MuEx}$~\cite{Aartsen_2014}) as shown in Fig.~\ref{fig:en_res}. At energies above 100 GeV there is a drop in the reconstructed neutrino energy. In the case of the $\textsc{PegLeg}$, this can be explained by the resulting muon leaving the IceCube detector volume. In addition the minimum-ionizing track assumption becomes inaccurate leading to an underestimation of the lost energy per unit of track length and an underestimation of the total neutrino energy. This bias, however, does not have a big impact on this analysis as we compare distributions using the same reconstruction.
\end{sloppypar}
 
After the new energy reconstruction is done, the analysis is performed on simulated data and the final BDT score threshold of the two selections are optimized so that they yield the best possible sensitivity (see Sec.~\ref{sec:method}). The optimized final cuts on the BDT scores are shown as a vertical lines on Fig.~\ref{fig:bdt}.
After this final cut the LE selection is left with 1,105 events, while the HE selection contains 7,414 events. Both the LE and HE selections have a $\sim90 - 95\%$ neutrino purity at this final level (see Figs.~\ref{fig:le_datamc},~\ref{fig:he_datamc} for the experimental data and Monte Carlo comparison after the likelihood minimization).

\section{The analysis method}
\label{sec:method}

\begin{figure*}
    \begin{minipage}{0.49\textwidth}
        \includegraphics[width=\textwidth]{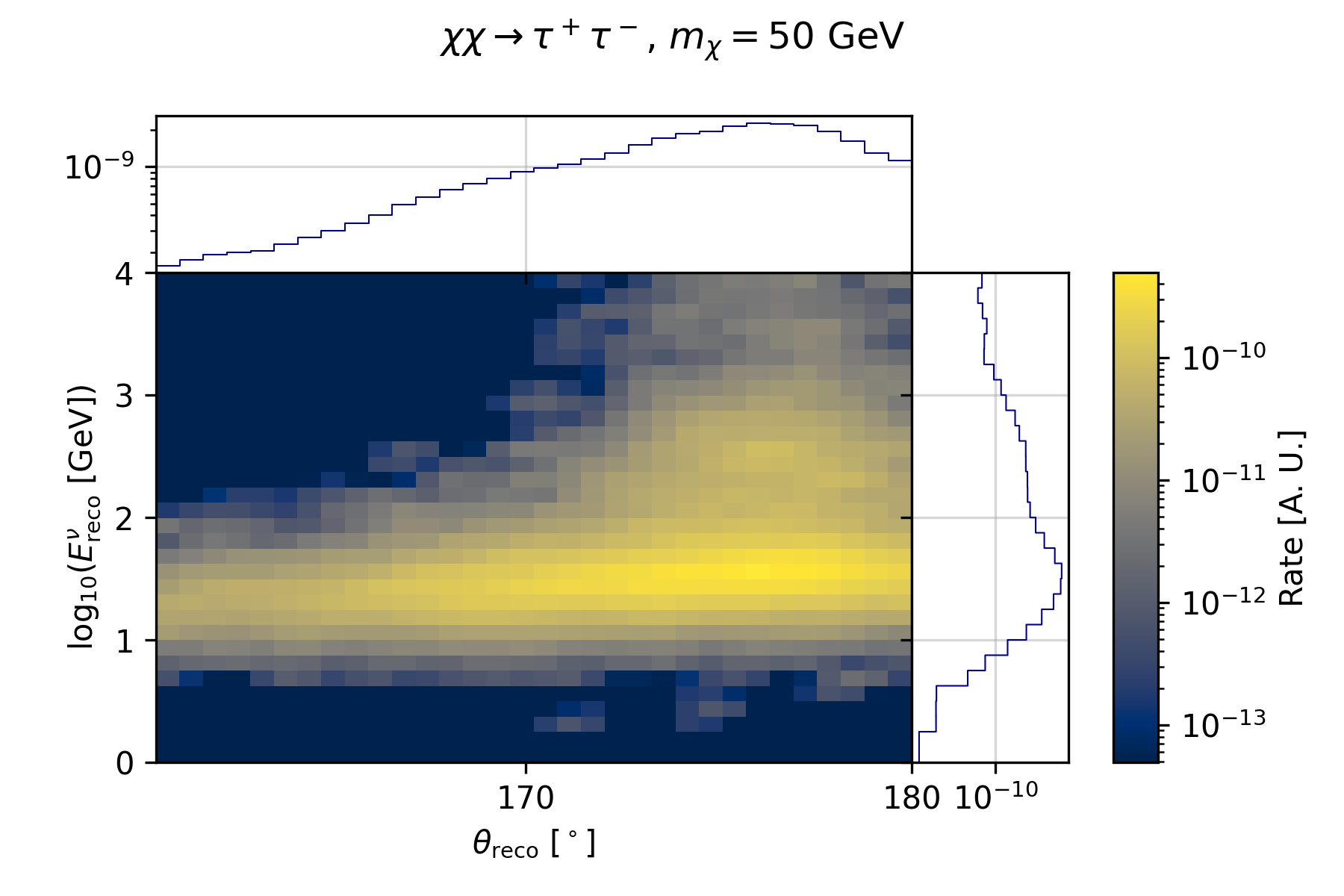}
    \end{minipage}
    \begin{minipage}{0.49\textwidth}
        \includegraphics[width=\textwidth]{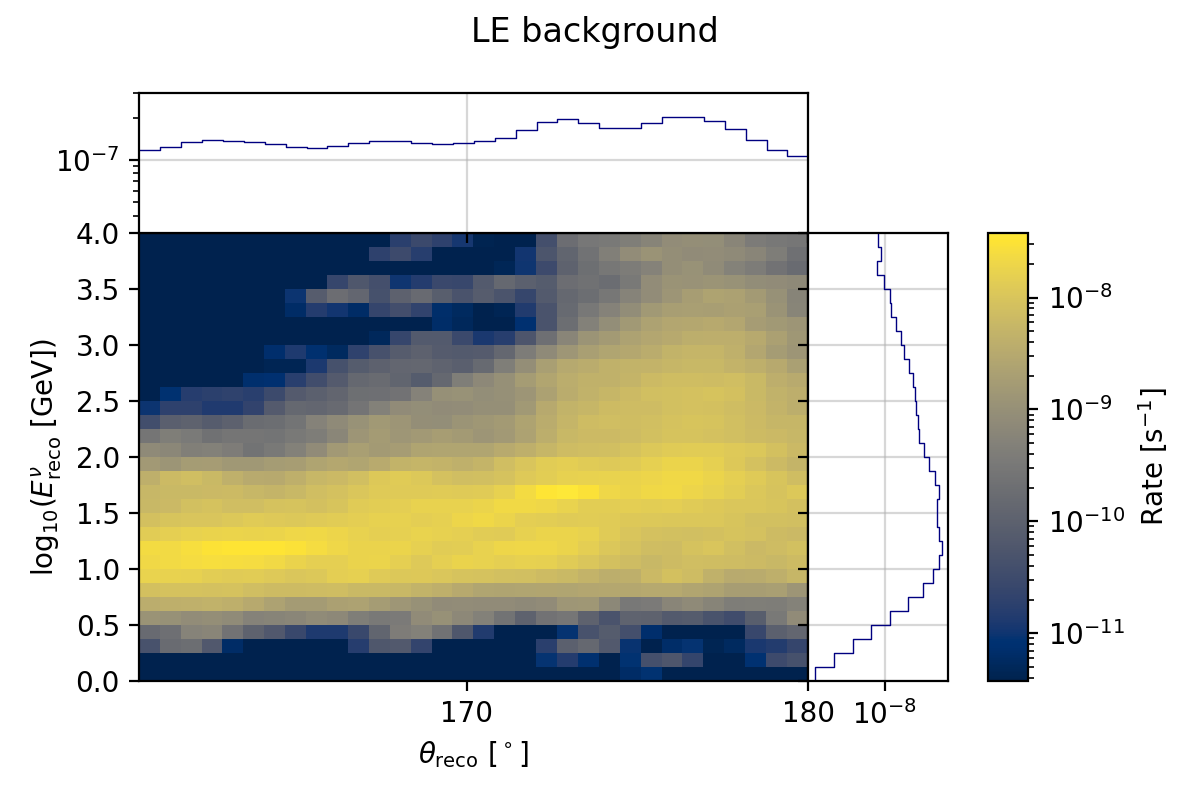}
    \end{minipage}
    \caption{Binned probability density functions for the LE analysis as a function of the reconstructed zenith angle and energy. The LE signal baseline and the atmospheric background are shown on the left and on the right, respectively. The signal normalization is set to an arbitrary number.}
    \label{fig:le_pdf}
\end{figure*}

\begin{figure*}
    \begin{minipage}{0.49\textwidth}
        \includegraphics[width=\textwidth]{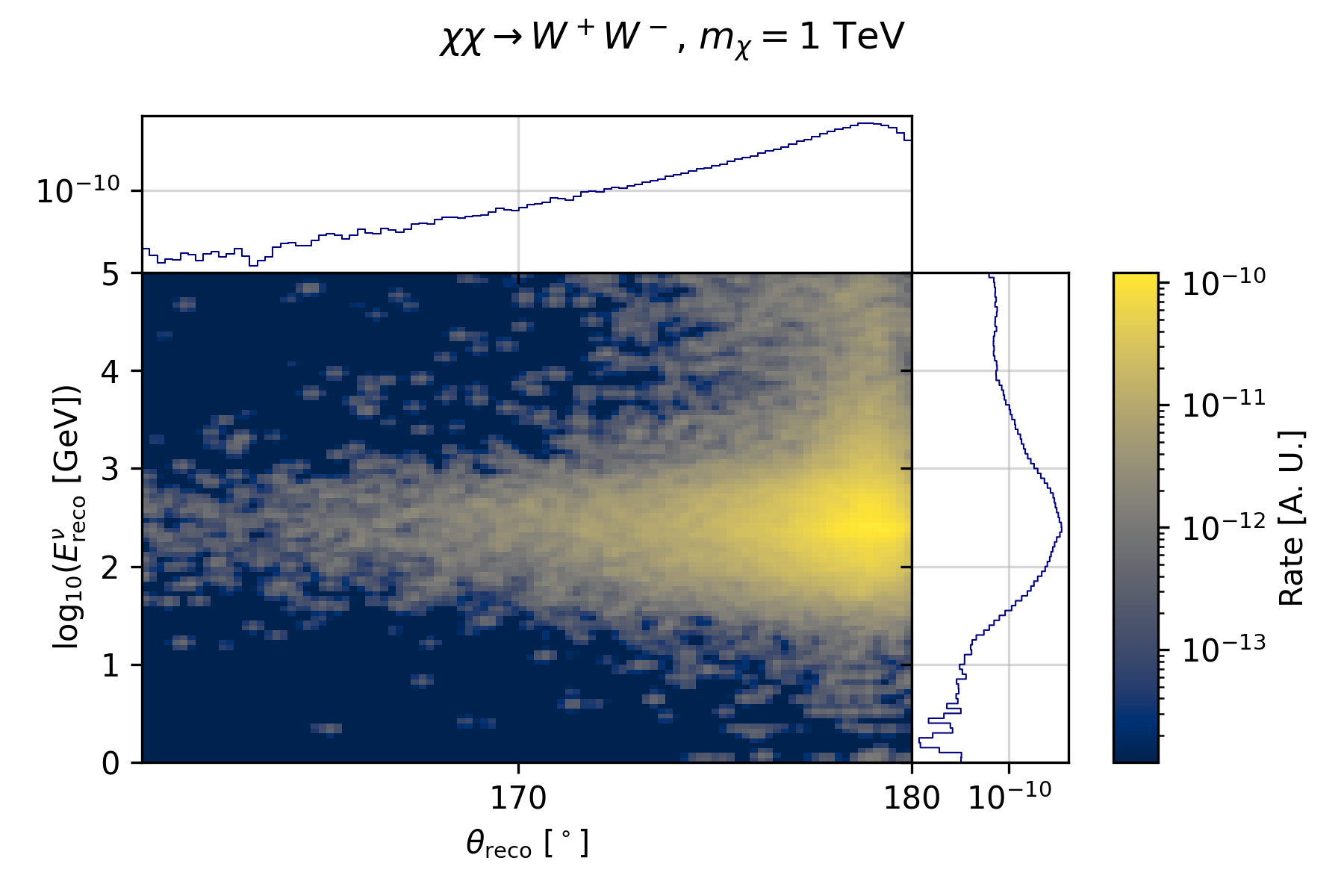}
    \end{minipage}
    \begin{minipage}{0.49\textwidth}
        \includegraphics[width=\textwidth]{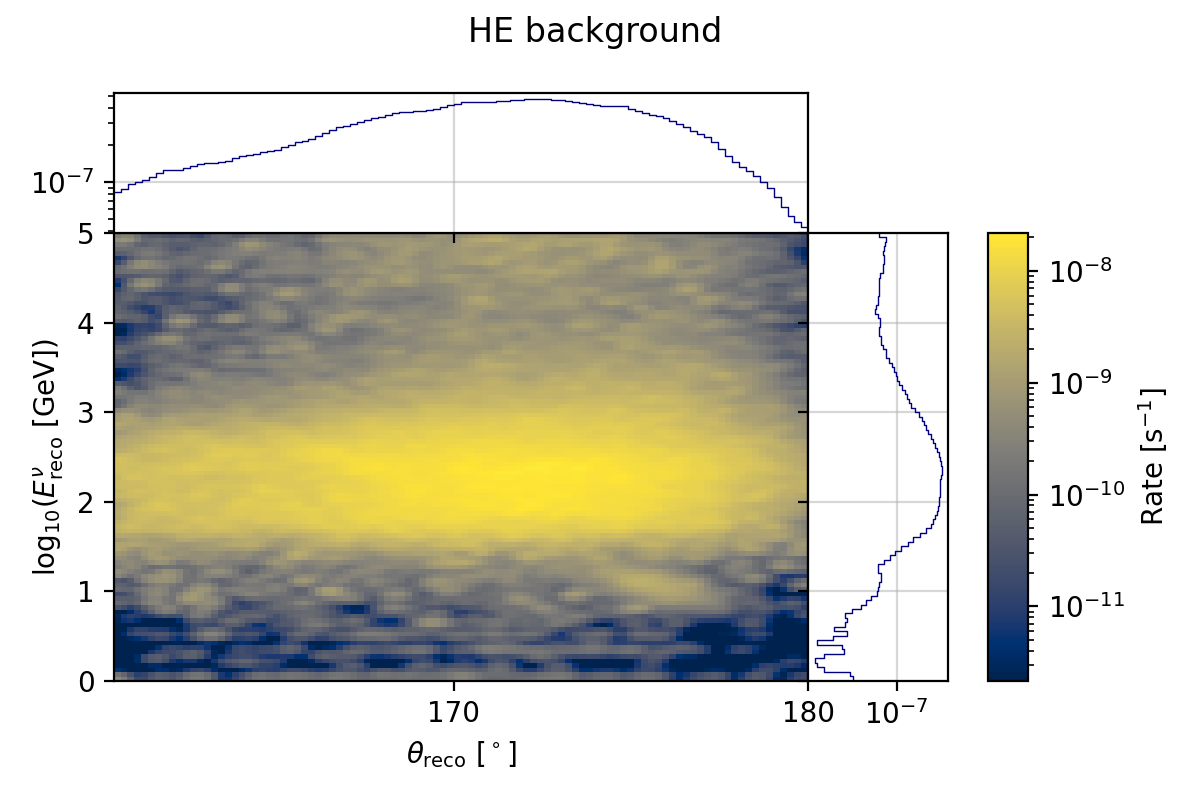}
    \end{minipage}
    \caption{Binned probability density functions for the HE analysis as a function of the reconstructed zenith angle and energy. The LE signal baseline and the atmospheric background are shown on the left and on the right, respectively. The signal normalization is set to an arbitrary number.}
    \label{fig:he_pdf}
\end{figure*}

We performed a likelihood ratio test for the analysis using a binned Poisson likelihood. We calculated probability density functions (PDFs) for both DM signal and background binned in a space of two observables: the reconstructed zenith angle, $\theta_\textrm{reco}$, and the reconstructed energy of the event, $\log_{10} E^\nu_\textrm{reco}$. These density functions were produced from simulations with limited statistics, and were smoothed with a \textit{Kernel Density Estimation} (KDE) technique to reduce the statistical fluctuations. The parameters of the KDE, such as its bandwidth, were optimized using a cross-validation method~\cite{scikit-learn}. For the LE analysis, we include events with energy in the range $1\;\textrm{GeV}<E^\nu_\textrm{reco}<10^4\;\textrm{GeV}$. We optimized the number of bins to be $32\times32$ for the PDF based on sensitivity studies. For the HE analysis, the energy range is $1\;\textrm{GeV}<E^\nu_\textrm{reco}<10^5\;\textrm{GeV}$ and the optimized bin grid is $100\times100$ bins. The different optimum bin sizes are due to the different statistics on both selections and underlying signal distributions. For both analyses, the zenith angle range is limited to $160^{\circ}<\theta_\textrm{reco}\leq180^{\circ}$. Two different background PDFs are calculated for the LE and HE selection, while the DM signal PDFs depend on the dark matter mass and annihilation channel. 
As is common in indirect searches of DM, each DM self-annihilation channel has been tested separately assuming a $100\%$ branching fraction. The PDFs for the signal, for one choice of DM mass and annihilation channel, and atmospheric background for LE and HE, are shown in Figs.~\ref{fig:le_pdf} and \ref{fig:he_pdf} respectively.

The expected number of events in the $i$th bin can be then written as a combination of the expected number of signal events plus the expected number of background events according to:

\begin{equation}
    \label{eq:model_base}
    \lambda_i = N_{\textrm{total}}(\xi S_i + (1-\xi)B_i),
\end{equation}

\noindent where $\xi$ is the total fraction of signal events in the full dataset, $N_{\textrm{total}}$ is the total number of events (in the LE or HE selection), and $S_i$ and $B_i$ are the signal and background PDFs respectively, representing the normalized fraction of signal and background events in the $i$th bin.
Given that the background includes multiple contributions, namely atmospheric neutrinos (including prompt) and muons, and astrophysical neutrinos, we decomposed the background distribution as follows:

\begin{equation}
    \label{eq:model}
B_i = \eta_0B_i^0 +\sum_{j=1}^{n-1}\left[\prod_{k=0}^{j-1}(1-\eta_k)\right]\eta_j B^j_i + \left[\prod_{k=0}^{n-1}(1-\eta_k)\right]B^n_i  
\end{equation} 

\noindent where $n$ is the number of different background components --conventional atmospheric, astrophysical neutrinos, prompt-- each contributing with a fraction that depends on the parameters $\overrightarrow{\eta}=(\eta_0, ...,\eta_{n-1})$. This nested formulation ensures a proper normalisation of the model as the sum of the parameters, ($\xi, \vec{\eta}$), is equal to $1$. Given the different description of neutrino interactions in \textsc{NuGen} and \textsc{Genie} (see Sec.~\ref{sec:icecube}), the atmospheric neutrino background was separated into \textsc{Genie} and \textsc{NuGen} components. Their relative normalizations are allowed to vary with a smooth transition in the range $100\textrm{\;GeV}<E_{\rm true}<200\textrm{\;GeV}$. 

With this formulation, where $n_i$ is the number of events observed in $i$th bin, we can build the Poisson likelihood $\mathcal{L}(\xi, \overrightarrow{\eta})$ as:

\begin{equation}
    \label{eq:likelihood}
    -\ln{\mathcal{L}(\xi,\overrightarrow{\eta})} = \sum_i^{N_{\textrm{bins}}}(-n_i\ln\lambda_i+\lambda_i + \ln n_i!),
\end{equation}

\noindent where $\xi$ is the physical parameter of interest and $\vec{\eta}$ are treated as nuisance parameters. The test statistic can be defined as~\cite{Cowan2011-xc}:

\begin{equation}
\label{eq:ts}
    t_\xi = 2\ln\frac{\mathcal{L}(\xi,\hat{\hat{\overrightarrow{\eta}}})}{\mathcal{L}(\hat{\xi},\hat{\overrightarrow{\eta}})} = 2\left[\ln\mathcal{L}(\xi,\hat{\hat{\overrightarrow{\eta}}})-\ln\mathcal{L}(\hat{\xi},\hat{\overrightarrow{\eta}})\right],
\end{equation}

\noindent where $\mathcal{L}(\xi,\hat{\hat{\overrightarrow{\eta}}})$ is the profiled maximum likelihood obtained by maximizing over $\vec{\eta}$ for any given value of $\xi$ and $\mathcal{L}(\hat{\xi},\hat{\overrightarrow{\eta}})$ is the global maximum likelihood. In the particular case of $\xi=0$, the discovery test statistic, $t_0$, can be used to evaluate the compatibility of the best fit parameter with the null hypothesis, $\mathcal{H}_0(\xi =0)$. After verifying that the test statistic asymptotically follows a $\chi^2$ distribution with one degree of freedom (see Wilks' theorem~\cite{Wilks:1938dza,Wilks2}), the pre-trial significance of the result in terms of the $z$-score --or number of standard deviations--, can be calculated as $z_0=\sqrt{t_0}$. When the significance is lower than $3\sigma$ ($z$-score = 3), we compute upper limits at $90\%$ confidence level (C.L.) by finding the value of $\xi$ such that the $t_\xi$ test statistic produces a significance of 10\% under the same $\mathcal{H}_1(\xi)$ hypothesis.
Before analyzing the data, predicted sensitivities are evaluated by calculating the median $90\%$ C.L. upper limits over 10,000 pseudo-experiments based on MC simulations. The choice of using the LE or HE dataset for a particular dark matter mass/annihilation channel is taken by selecting the one that yields the best sensitivity. Based on this criterion, DM particle masses $m_{\chi}\leq100\;\textrm{GeV}$ are analysed using the LE dataset, while for $m_{\chi}>100\;\textrm{GeV}$ the HE dataset is applied. As previously mentioned, the best achievable sensitivity is also used to determine the optimum score cut in the BDTs for both the LE and HE selection. 

\paragraph{Constraints.---}
\begin{sloppypar}
The only constraint imposed on the physical parameter, $\xi$, is that it must be positive, $\xi \ge 0$. For the nuisance parameters, $\vec{\eta}$, two constraints were added based on previous knowledge about the normalization of the different background components. This information was incorporated into the likelihood formalism in the form of Gaussian constraint terms. Specifically, a constraint on the astrophysical contribution was applied to limit its normalization within the $1\sigma$ band of the most recent IceCube measurement~\cite{IceCubeDiffuse:2021uhz}. Additionally, a second constraint was introduced to ensure that the relative contributions of \textsc{Genie} and \textsc{NuGen} do not deviate more than $20\%$, which conservatively encompasses the observed relative difference between the two simulations of about $10\%$.   
\end{sloppypar}

\paragraph{Systematic uncertainties.---}
Several sources of systematic uncertainties are included in this analysis. Some of them are included as nuisance parameters in the likelihood formalism, such as the individual normalization of the different background components. Others, due to limitation in the simulation of such systematic effects, are considered as discrete parameters. In these cases, the likelihood is minimized using the different systematic variations and the minimum of all these likelihoods is selected. The sources of systematic uncertainty in this analysis can be classified in three groups: 

\begin{itemize}
    \item \textbf{Uncertainties in the propagation of light in the Antarctic ice.} One particular uncertainty in the propagation of light through the ice, is the optical properties of the medium. In particular, for vertical tracks which affects this analysis the most, the modeling of the {\it hole ice} is of special importance. This is the ice which froze around the DOMs after the hot water drilling used to put the strings in place. Hole ice has been observed to have different optical properties than the common South Pole ice or bulk ice. The hole ice optical properties can be modeled using two parameters~\cite{Leuermann:2018oxl}. The first affects the acceptance of incoming photons depending on their incoming direction, and the second influences vertical up-going photons. Discrete variations of these parameters are used in simulation and included in the systematic uncertainty treatment. The uncertainty in the exact position of the bedrock at the South Pole is an unknown that could also influence the neutrino interaction and photon propagation. For neutrinos incoming from the direction of the center of the Earth, the impact of the bedrock position is smaller than typical statistical fluctuations, of $\sim1\%$, hence this uncertainty is not considered in this analysis. 
    
    \item \textbf{Uncertainties in the detector response.} The main driver in the uncertainties of the detector response is the DOM efficiency, which is influenced by the photomultiplier tube efficiency, hardware effects, as well as the surrounding ice. Discrete variations in efficiency ($\pm 10$\%) are considered in this analysis. These variation affect event detection rate as well as the shape of PDFs. 

    \item \textbf{Uncertainties in the dark matter halo model.} To calculate the capture rate we adopted the \textit{Standard Halo Model}, in which the velocity distribution of DM particles in the Galactic halo is assumed to be a truncated Maxwellian distribution with a dispersion of 270 km/s and an escape velocity of 544 km/s. The local DM density is taken to be 0.3 GeV/cm$^3$. Recent astronomical data as well as numerical simulations of structure formation in cold DM indicate that there may be deviations from these assumptions, however for consistency with other experimental analyses we continue with these recommended parameters \cite{Baxter:2021pqo}. Note that we cannot adopt the halo model-independent formalism \cite{Ferrer_2015} that was used in a previous IceCube analysis of neutrinos from  the Sun \cite{IceCube&PICO:2019lus}, since DM captured by the Earth has not yet equilibrated.

    \item \textbf{Physics uncertainties.} Finally uncertainties in the physical quantities, such as the atmospheric, astrophysics fluxes, and oscillation parameters were also considered. Two alternative atmospheric models were tested in addition to the nominal \textit{Honda}~\cite{Honda2006:PhysRevD.75.043006,Honda2014:PhysRevD.92.023004} model: the \textit{Corsika} model~\cite{Schoneberg:2016rkg}, and the \textit{Bartol}~\cite{Barr:PhysRevD.70.023006} model. To cope with the different response of the two neutrino generators used in this analysis, \textsc{Genie} and \textsc{NuGen}, in the region around \mbox{$\sim 100$ GeV} an additional systematic parameter is introduced. This parameter allows for the two components to vary relative to one another by gradually increasing \textsc{NuGen} events in the region $100\textrm{\;GeV}<E_{\rm true}<200\textrm{\;GeV}$ from 0\% to 100 \% while \textsc{Genie} contribution evolves in the opposite way. The expected contribution of astrophysical neutrinos is expected to be less than $1\%$ of the total number of events in this analysis. However the normalization of this contribution is also included as a nuisance parameter. Finally, propagation through the Earth enhances the neutrino oscillation probability causing $\nu_\mu$ disappearance and $\nu_\tau$ appearance effects. Several oscillation systematic variations parameters were used, including the IceCube fit values from~\cite{OscIce:2017}, an inverted mass order set from \cite{Zyla:2020zbs} and a set with a $\delta_{CP}$ phase, with the $\delta_{CP}$ value taken from \cite{Zyla:2020zbs}.

\end{itemize}

\section{Results}
\label{sec:results}

\begin{figure*}
    \begin{minipage}{0.49\textwidth}
        \includegraphics[width=\textwidth]{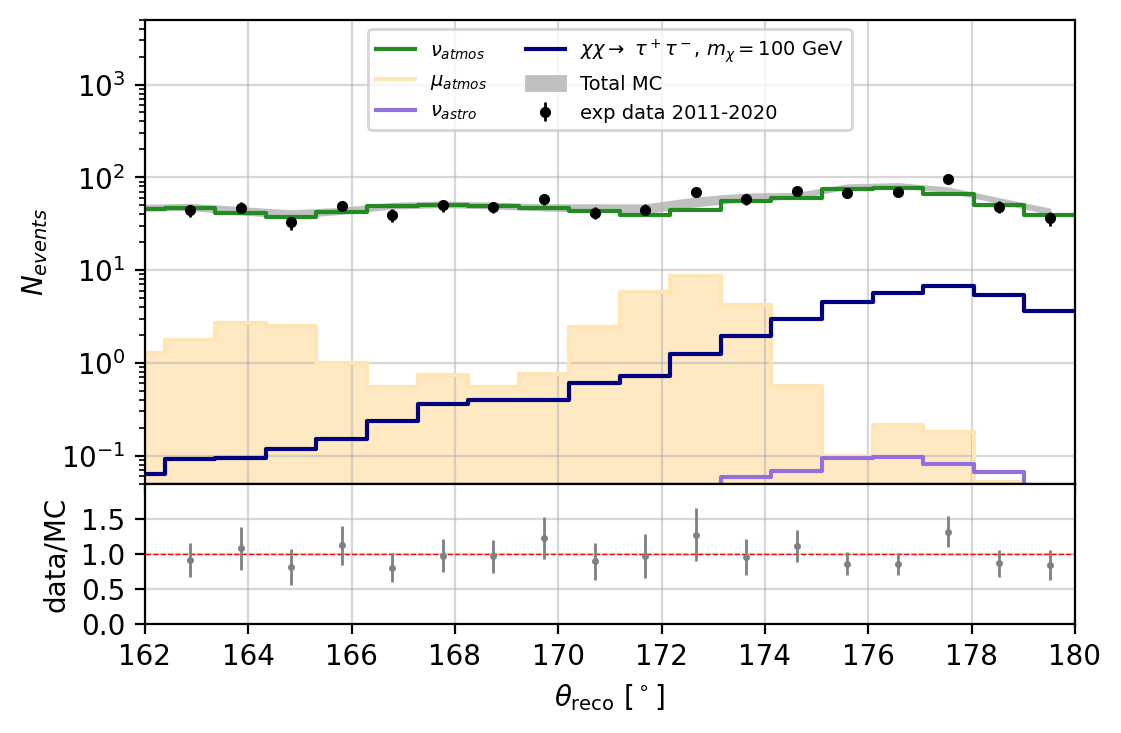}
    \end{minipage}
    \begin{minipage}{0.49\textwidth}
        \includegraphics[width=\textwidth]{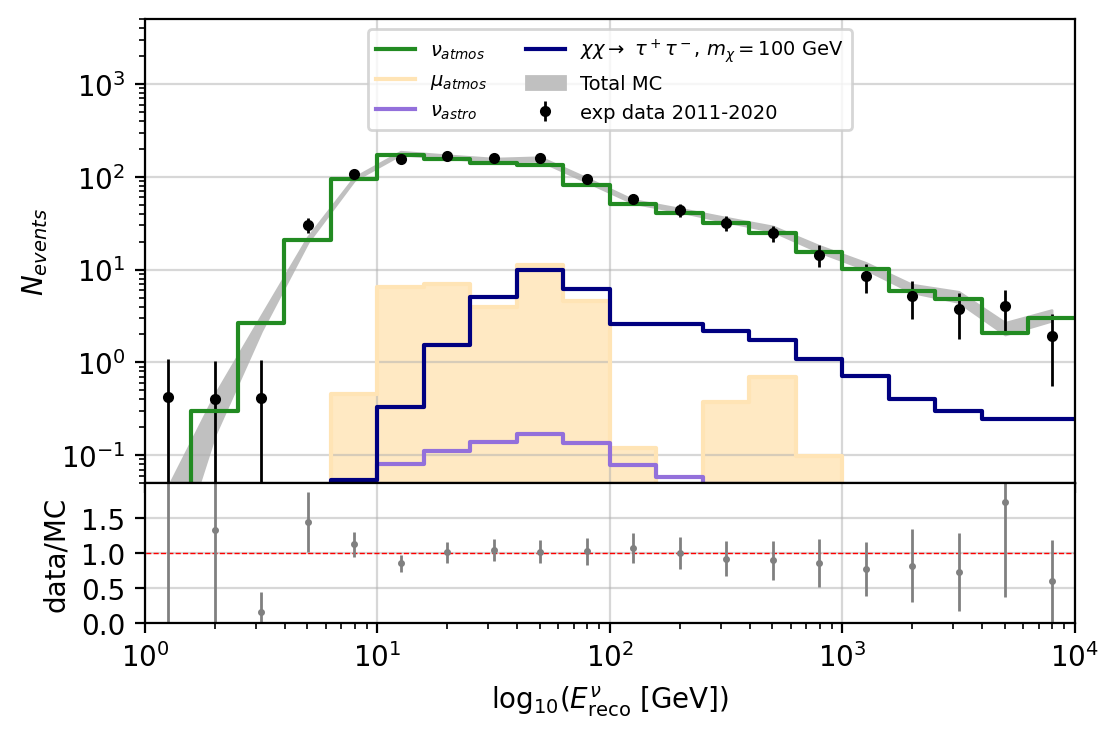}
    \end{minipage}
    \caption{Data and MonteCarlo distributions as a function of the reconstructed zenith angle (left) and energy (right) for the LE analysis for the best fit value in the $\chi\chi \rightarrow \tau^+\tau^-$ channel for a dark matter mass of $m_\chi$ = 100 GeV. Atmospheric neutrinos include prompt and the individual contribution of \textsc{Genie} and \textsc{NuGen}.}
    \label{fig:le_datamc}
\end{figure*}

\begin{figure*}
    \begin{minipage}{0.49\textwidth}
        \includegraphics[width=\textwidth]{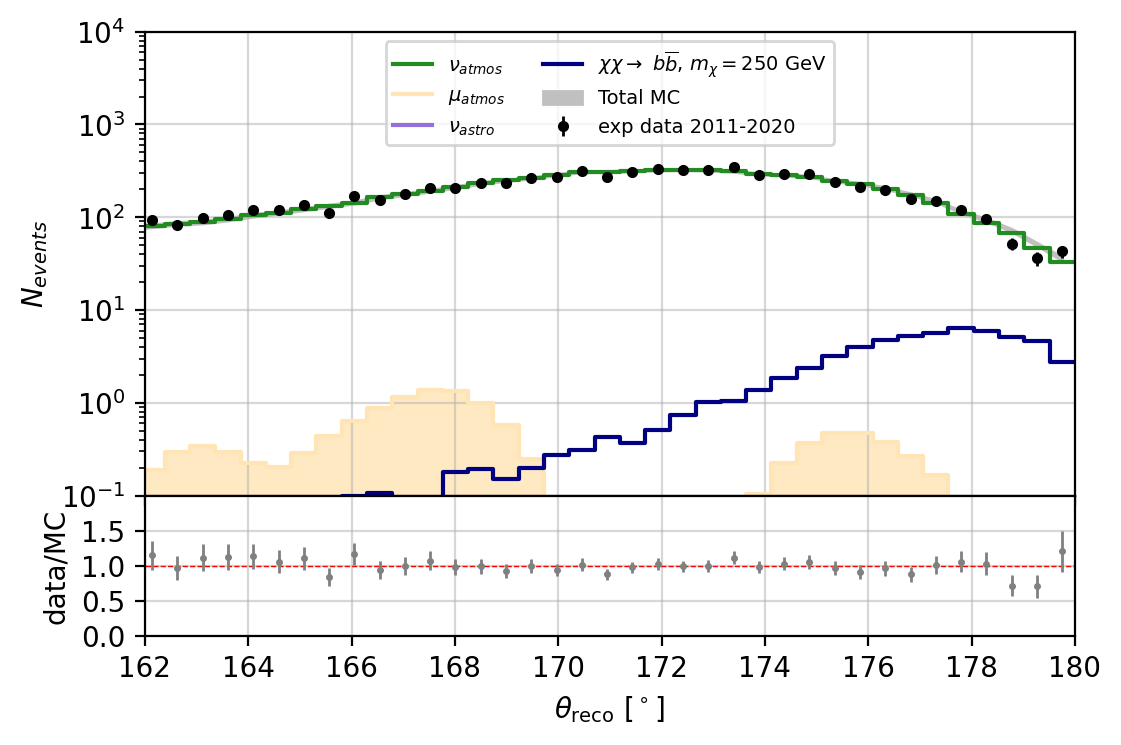}
    \end{minipage}
    \begin{minipage}{0.49\textwidth}
        \includegraphics[width=\textwidth]{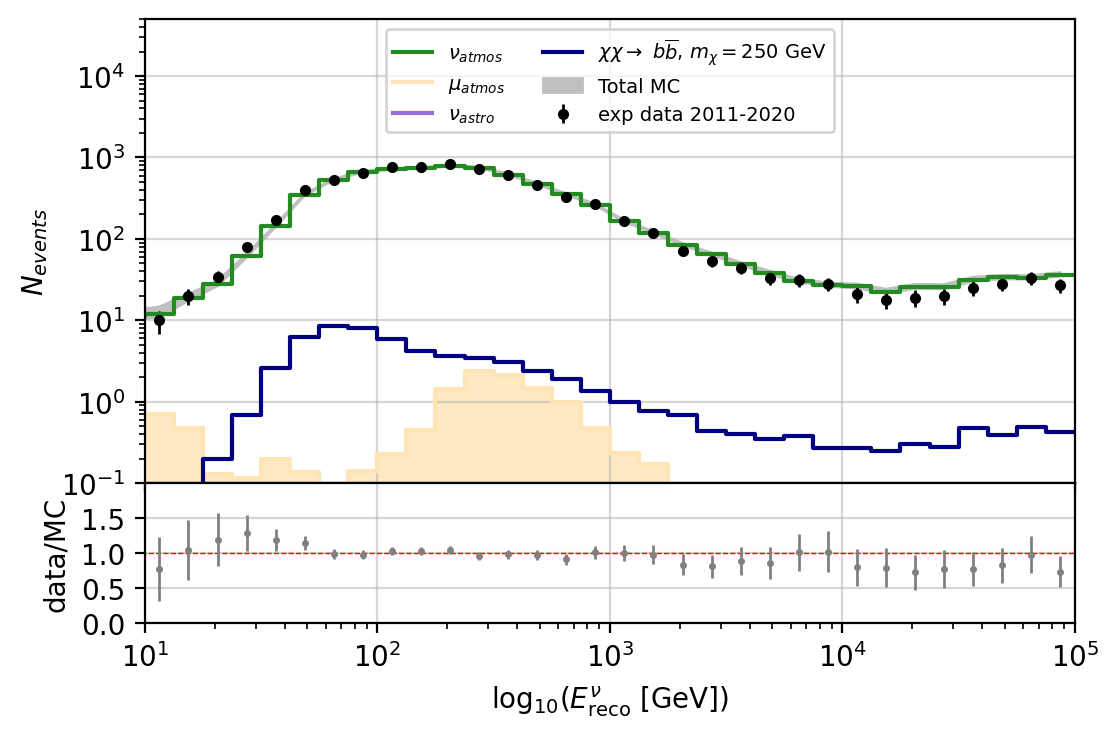}
    \end{minipage}
    \caption{Data and MonteCarlo distributions as a function of the reconstructed zenith angle (left) and energy (right) for the HE analysis for the best fit value in the $\chi\chi \rightarrow b\bar{b}$ channel for a dark matter mass of $m_\chi$ = 250 GeV. Atmospheric neutrinos include prompt and the individual contribution of \textsc{Genie} and \textsc{NuGen}.}
    \label{fig:he_datamc}
\end{figure*}

After maximizing the likelihood for each of the masses and channels listed in Tab.~\ref{table:wimpsim} we found no significant excess of a dark matter signal coming from the center of the Earth. The most significant pre-trial $p$-value, $1.94\sigma$, was found in the HE analysis for the channel $\chi\chi\rightarrow b\bar{b}$ and  mass \mbox{$m_{\chi}=250$ GeV}. The trial factor correction, accounting for all the masses and decay channels, yields a post-trial significance of $1.06\sigma$.  The rest of nuisance parameters were found to be within the constrains. The final data and Monte Carlo comparison plots found in the two best fit scenarios for the LE sample ($\chi\chi \rightarrow \tau^+\tau^-$, $m_\chi$ = 100 GeV) and for the HE sample ($\chi\chi \rightarrow b\bar{b}$ at $m_\chi$ = 250 GeV) can be seen in Figs.~\ref{fig:le_datamc} and~\ref{fig:he_datamc} respectively.

Given the null result, we provide upper limits on the spin-independent DM-nucleon cross section $\sigma_{\chi N}^\textrm{SI}$. In particular, we can estimate an upper limit on the volumetric flux of muons from neutrino interactions as:

\begin{equation}
        \label{eq:vol_flux}
        \Gamma_{\nu\rightarrow\mu}^{\textrm{90\%C.L.}} = \frac{\xi^{90\%\textrm{C.L.}}N_{\textrm{total}}}{V_{\textrm{eff}}\cdot t_{\textrm{livetime}}},
    \end{equation}

\noindent where $t_{\textrm{livetime}}$ corresponds to the total analyzed livetime (3,266 days), $V_{\textrm{eff}}$ is the effective volume of the detector, and the quantity $\xi^{90\%\textrm{C.L.}}N_{\textrm{total}}$ represents the upper limit on the number of signal neutrinos in the sample. 

\begin{figure}
    \includegraphics[width=.49\textwidth]{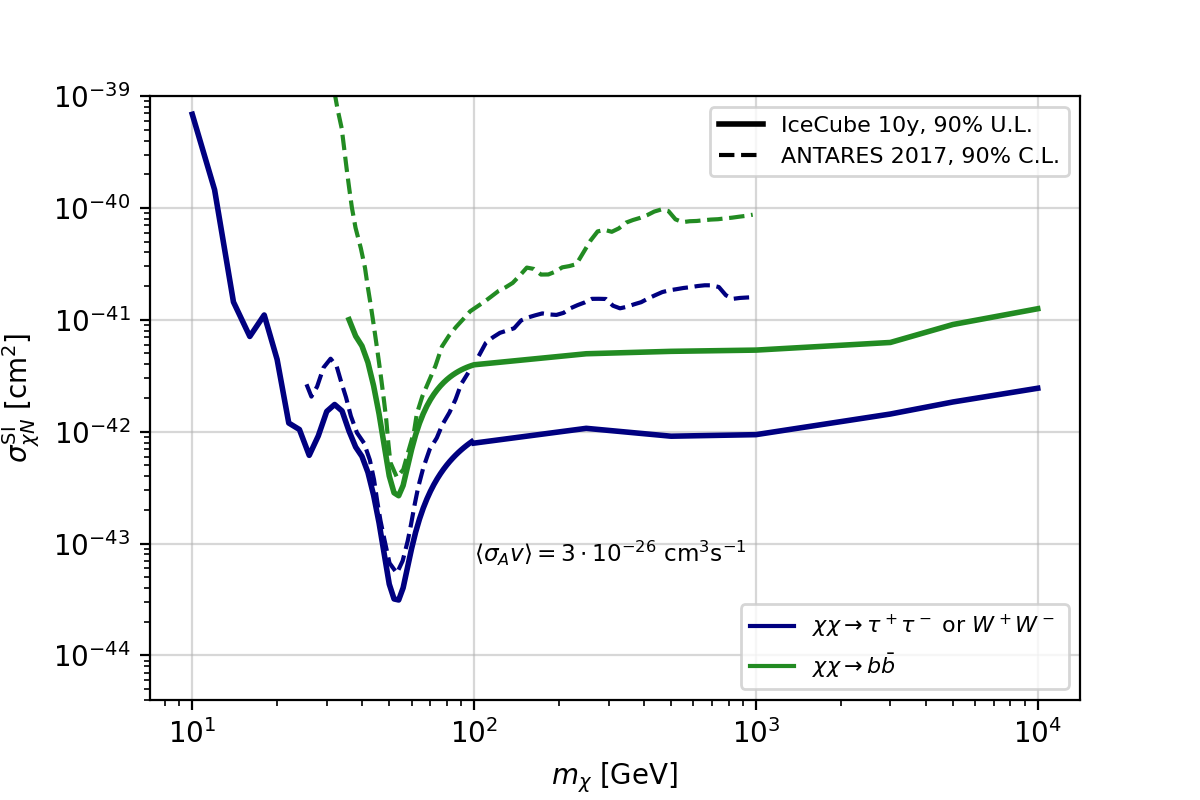}
    \caption{Spin-independent DM-nucleon cross section $90\%$ C.L. upper limits. Limits in bold are from this analysis, and dashed for ANTARES. The colour code identifies the annihilation channels, blue for annihilation into $\tau^+\tau^-$ ($W^+W^-$ from 80.4 GeV) and green for $b\bar{b}$.}
    \label{fig:xsec_limits}
\end{figure}

The $90\%$ C.L. upper limits for $\sigma_{\chi N}^{\rm SI}$ are presented in Fig.~\ref{fig:xsec_limits}, compared to results from the  ANTARES neutrino telescope~\cite{ANTARES_Earth:2016}.  
\begin{sloppypar}
    
\end{sloppypar}

\begin{figure}
    \includegraphics[width=.49\textwidth]{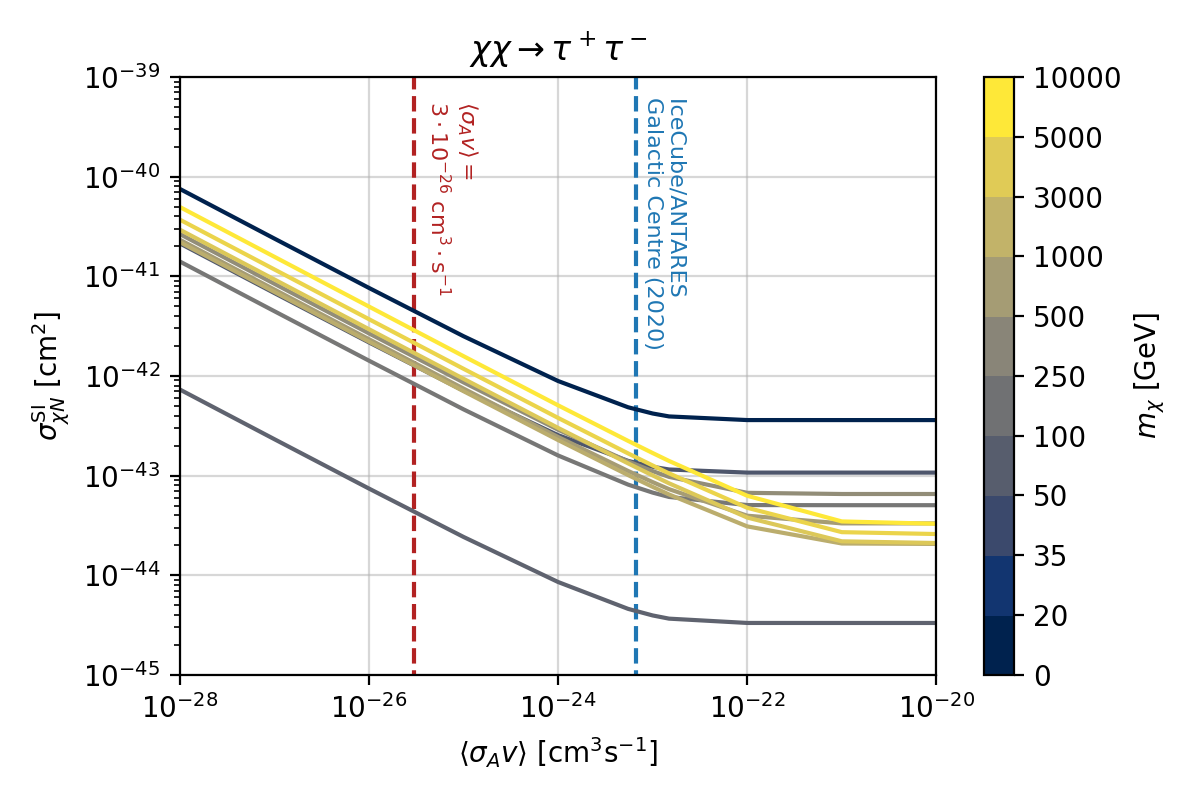}
    \includegraphics[width=.49\textwidth]{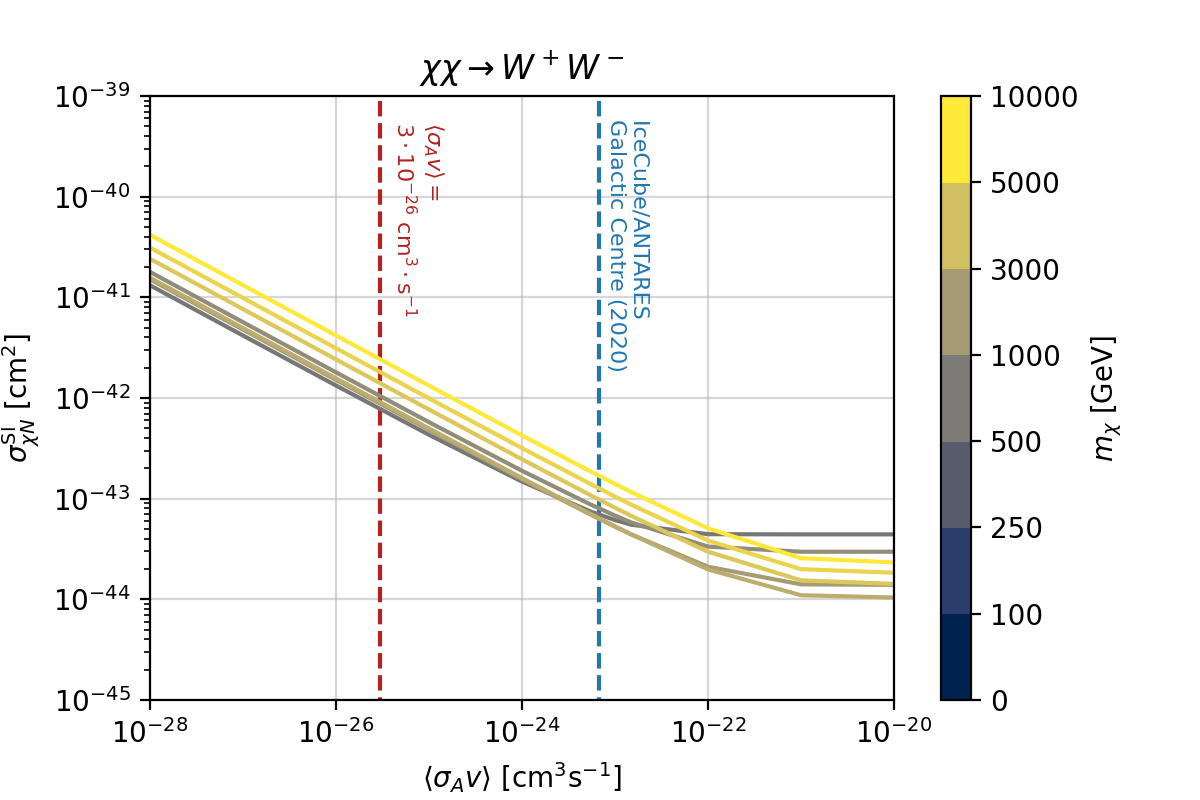}
    \includegraphics[width=.49\textwidth]{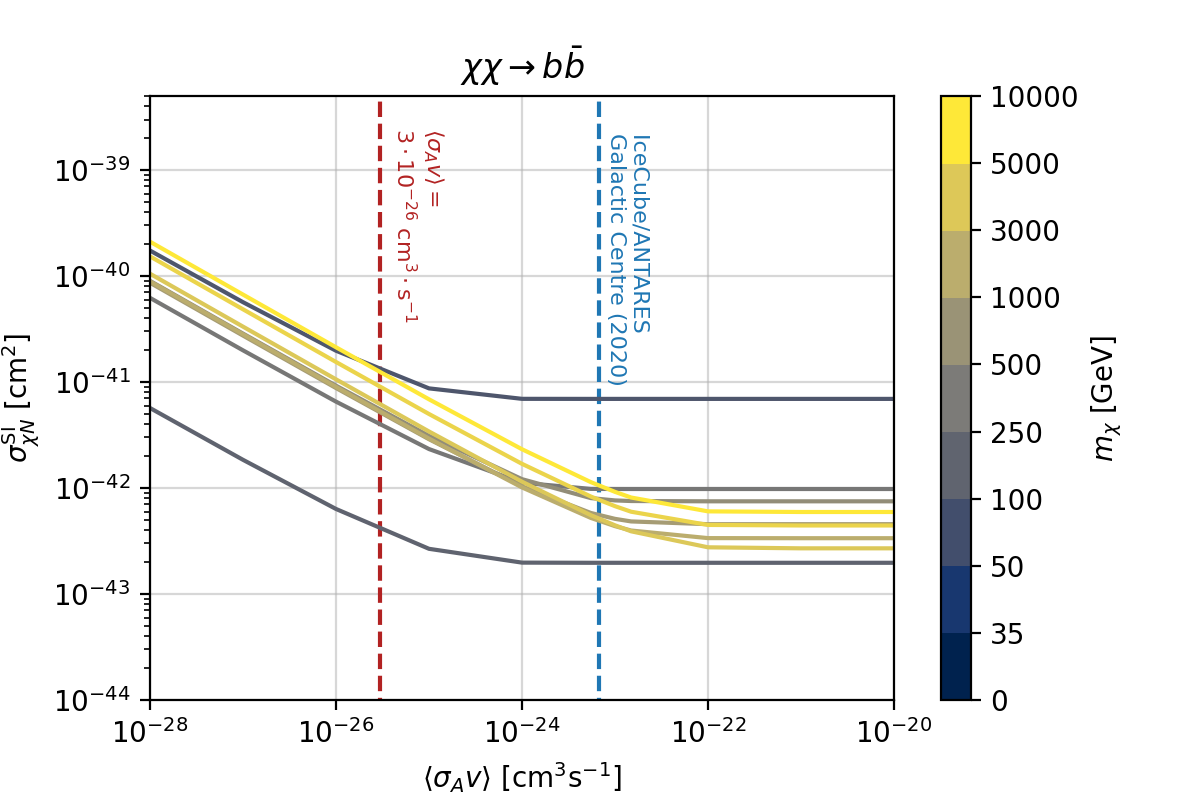}
    \caption{Upper limits at 90\% C.L. on the $\sigma_{\chi N}^{\rm SI}$ as a function of the $\langle\sigma_{\textrm{A}}v\rangle$ (on the $x$-axis) for the $\chi\chi\rightarrow \tau^+\tau^-$ (top), $\chi\chi\rightarrow W^+W^-$ (middle), and $\chi\chi\rightarrow b\bar{b}$ (bottom) annihilation channels. The color scale indicates DM particle mass. The two vertical lines indicate the canonical thermal annihilation cross section (dotted red) and the averaged combined IceCube/ANTARES limit (90\% C. L.) on the annihilation cross section from the Galactic Halo (dotted blue).}
    \label{fig:xsecvs}
\end{figure}

A more general representation of the upper limits on the scattering cross section, without any assumption regarding the annihilation cross section, can be done in the $\langle\sigma_{\textrm{A}}v\rangle - \sigma_{\chi N}^{\textrm {SI}}$ plane as illustrated in Fig.~\ref{fig:xsecvs} for different channels and masses. As can be seen, limits on $\sigma_{\chi N}^{\textrm {SI}}$ are stronger for larger values of $\langle\sigma_{\textrm{A}}v\rangle$. A large enough value of $\langle\sigma_{\textrm {A}}v\rangle$ will ensure equilibrium between capture and self-annihilation rate even in the Earth. This is visible in the figures as a horizontal line where for any value of $\langle\sigma_{\textrm{A}}v\rangle$ the limit on $\sigma_{\chi N}^{\textrm {SI}}$ remains constant. As a reference, the vertical red dotted line indicates the canonical assumption of a thermal WIMP self-annihilation cross section, while the vertical blue dotted line is the mass averaged combined limit from IceCube and ANTARES on the WIMP self-annihilation cross section from observations of the Galactic center~\cite{CombinedGC:2020}.

\section{Discussion and outlook}
\label{sec:discu}

The upper limits set in this analysis represent an improvement over the first IceCube search for dark matter annihilation at the center of the Earth~\cite{EarthIce:2016} by a factor of $>3$ for any mass or channel analyzed, and in particular by an order of magnitude for $m_{\chi}>100$ GeV. Figure~\ref{fig:xsec_limits} shows how, for such heavy DM, this analysis gives the best limits for all self-annihilation channels. The major improvement in sensitivity for $m_{\chi}>100$ GeV is ascribable to the use of the energy as an observable in the present analysis. 

\begin{figure}
    \includegraphics[width=.49\textwidth]{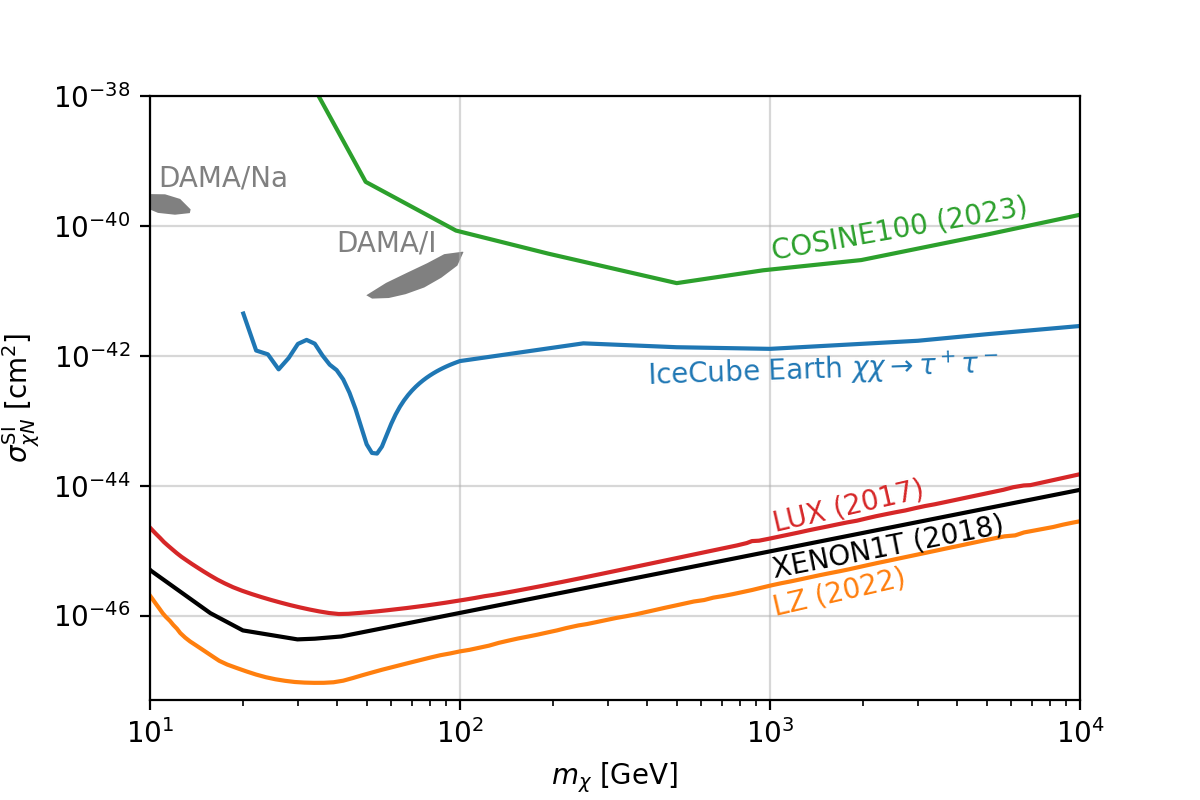}
    \caption{Upper limits on the spin independent nucleon dark matter cross section for this analysis (blue) for the $\chi\chi\rightarrow \tau^+\tau^-$ annihilation channel compared to direct detection upper limits from the crystal experiments DAMA/LIBRA \cite{DAMA:2018} (grey areas) and COSINE100 \cite{COSINE100:2019} (green). Also shown are the results from LUX~\cite{LUX:2017} (red), XENON1T~\cite{XENON3:2019} (black), and first results from LUX-ZEPLIN (LZ) experiment~\cite{LZ:2022} (orange).}
    \label{fig:vs_dir}
\end{figure}

\begin{sloppypar}   
In Fig.~\ref{fig:vs_dir}, we show a comparison with results from direct detection experiments. The upper limits from this analysis are lower than those of crystal experiments like COSINE100~\cite{COSINE100:2019}. Although limits from liquid xenon detectors, such as XENON1T~\cite{XENON3:2019}, LUX~\cite{LUX:2017}, or the LUX-Zeplin experiment~\cite{LZ:2022} are an order of magnitude better, we consider our results complementary in view of the uncertain DM velocity distribution. In fact, DM particles are more likely to be captured when they have relatively low velocity while in direct detection experiments, to produce a detectable nuclear recoil, high-velocity particles are needed. Therefore, the two types of searches probe different regions of the DM velocity distribution.
\end{sloppypar}

As the results from this analysis are based on the number of signal events, they can be simply recast to test different DM models. See for example Ref.~\cite{Renzi:2022}, where limits on the coupling constant of the effective field theory of dark matter~\cite{Catena_2017} were computed.

Opening the selection to all-neutrino-flavors will improve the sensitivity of this analysis. In addition, with the forthcoming IceCube Upgrade~\cite{Ishihara:2019aao,Baur:2019}, there is a significant potential for improvement in the low energy region. 

\section{Conclusion}
\label{sec:conclu}

We conducted a search for dark matter annihilating at center of the Earth with IceCube. The peculiar position of the source required extensive use of Monte Carlo simulations and the development of a dedicated event selection, which was split into a low-energy and a high-energy selection, both of which achieve a $\sim90\%$ neutrino purity. We found no excess over background for the neutrino flux from the center of the Earth. The most significant result is for DM mass $m_{\chi}=250$ GeV and $\chi\chi\rightarrow b\bar{b}$ annihilation channel, which had a post-trial significance of $1.06\sigma$. Therefore, we place upper limits on the spin-independent DM-nucleon cross section which are currently the best limits set by a neutrino telescope for $m_{\chi}>100$ GeV. A significant improvement in sensitivity is expected for lower masses following the installation of the upcoming IceCube Upgrade.

\begin{acknowledgements}
The IceCube collaboration acknowledges the significant contributions to this manuscript from J. A. Aguilar and G. Renzi.
The authors gratefully acknowledge the support from the following agencies and institutions:
USA {\textendash} U.S. National Science Foundation-Office of Polar Programs,
U.S. National Science Foundation-Physics Division,
U.S. National Science Foundation-EPSCoR,
U.S. National Science Foundation-Office of Advanced Cyberinfrastructure,
Wisconsin Alumni Research Foundation,
Center for High Throughput Computing (CHTC) at the University of Wisconsin{\textendash}Madison,
Open Science Grid (OSG),
Partnership to Advance Throughput Computing (PATh),
Advanced Cyberinfrastructure Coordination Ecosystem: Services {\&} Support (ACCESS),
Frontera computing project at the Texas Advanced Computing Center,
U.S. Department of Energy-National Energy Research Scientific Computing Center,
Particle astrophysics research computing center at the University of Maryland,
Institute for Cyber-Enabled Research at Michigan State University,
Astroparticle physics computational facility at Marquette University,
NVIDIA Corporation,
and Google Cloud Platform;
Belgium {\textendash} Funds for Scientific Research (FRS-FNRS and FWO),
FWO Odysseus and Big Science programmes,
and Belgian Federal Science Policy Office (Belspo);
Germany {\textendash} Bundesministerium f{\"u}r Bildung und Forschung (BMBF),
Deutsche Forschungsgemeinschaft (DFG),
Helmholtz Alliance for Astroparticle Physics (HAP),
Initiative and Networking Fund of the Helmholtz Association,
Deutsches Elektronen Synchrotron (DESY),
and High Performance Computing cluster of the RWTH Aachen;
Sweden {\textendash} Swedish Research Council,
Swedish Polar Research Secretariat,
Swedish National Infrastructure for Computing (SNIC),
and Knut and Alice Wallenberg Foundation;
European Union {\textendash} EGI Advanced Computing for research;
Australia {\textendash} Australian Research Council;
Canada {\textendash} Natural Sciences and Engineering Research Council of Canada,
Calcul Qu{\'e}bec, Compute Ontario, Canada Foundation for Innovation, WestGrid, and Digital Research Alliance of Canada;
Denmark {\textendash} Villum Fonden, Carlsberg Foundation, and European Commission;
New Zealand {\textendash} Marsden Fund;
Japan {\textendash} Japan Society for Promotion of Science (JSPS)
and Institute for Global Prominent Research (IGPR) of Chiba University;
Korea {\textendash} National Research Foundation of Korea (NRF);
Switzerland {\textendash} Swiss National Science Foundation (SNSF).

\end{acknowledgements}

\bibliographystyle{spphys}
\bibliography{references}
%
%

\end{document}